\documentclass[aps,pre,twocolumn,groupedaddress]{revtex4-1}

\usepackage{amsmath}
\usepackage{amssymb}
\usepackage{graphicx}
\usepackage{dcolumn}
\usepackage{bm}
\usepackage[english]{babel}
\usepackage{enumitem}
\usepackage{soul}
\usepackage{epstopdf}
\usepackage{wasysym}

\usepackage{pifont}

\usepackage{blindtext} 
\usepackage{tcolorbox} 

\usepackage{hyperref}         
\hypersetup{colorlinks=true}  

\makeatletter
\let\reftagform@=\tagform@
\def\tagform@#1{\maketag@@@{\color{blue}(\ignorespaces{#1}\unskip\@@italiccorr)}}
\renewcommand{\eqref}[1]{\textup{\reftagform@{\ref{#1}}}}
\makeatother

\usepackage{titlesec}
\usepackage{xcolor,xpatch}
\colorlet{sectitlecolor}{blue} 
\colorlet{sectboxcolor}{white!30} 
\colorlet{secnumcolor}{blue} 
\titleformat{\section} {\normalfont\large\color{sectitlecolor}}{\colorbox{sectboxcolor}{\textcolor{secnumcolor}{\thesection}}}{1em}{} 
\titleformat{\subsection} {\normalfont\normalsize\color{sectitlecolor}}{\colorbox{sectboxcolor}{\textcolor{secnumcolor}{\thesubsection}}}{1em}{} 
\titleformat{\subsubsection} {\normalfont\normalsize\color{sectitlecolor}}{\colorbox{sectboxcolor}{\textcolor{secnumcolor}{\thesubsection}}}{1em}{} 

\usepackage{color} 
\definecolor{PaleYellow1}{rgb}{1.0,1.0,0.5}
\definecolor{Yellow1}{rgb}{1.0,1.0,0.5}
\definecolor{LinkColor}{rgb}{0.0,0.45,0.0} 
\definecolor{Grey0}{rgb}{0.95,0.95,0.95} 
\definecolor{Grey1}{rgb}{0.92,0.92,0.92} 
\definecolor{Grey2}{rgb}{0.4,0.4,0.4} 
\definecolor{Green1}{rgb}{0.4,0.9,0.4}
\definecolor{Orange1}{rgb}{1.0,0.7,0.05}
\definecolor{DarkBlue}{rgb}{0,0.08,0.45}
\definecolor{DarkRed}{rgb}{0.75,0.08,0.0}
\definecolor{BrightGrey}{rgb}{0.4,0.4,0.4}
\definecolor{BrightGreen}{rgb}{0.1,0.8,0.1}
\definecolor{Orange}{rgb}{1.0,0.5,0.01}
\definecolor{BlueGreen}{RGB}{12,201,179}
\definecolor{kbficolor}{RGB}{171,83,83}
\definecolor{etiscolor}{RGB}{81,106,156}

\newcommand{\B}{\color{blue}}

\hypersetup{
    colorlinks=true,
    linkcolor=blue,                  
    filecolor=red,   
    urlcolor=blue,                   
    citecolor=LinkColor,             
    anchorcolor=Grey2,
    pdftitle={M.Patriarca - Statistical correlations in the oscillator model of quantum dissipative systems}, 
    bookmarks=true,
    pdfpagemode=FullScreen,
}

\newcommand{\cRef}[1]{{\color{LinkColor} \cite{#1}}} 

\makeatletter
\xpatchcmd{\@lbibitem}
 {\item[\hfil}
 {\item[\hfil\color{LinkColor}}
 {}{}
\makeatother 

\newcommand{\req}[1]{{\color{blue}({\ref{#1}})}} 
\newcommand{\eq}[1]{Eq.~{\color{blue}(\ref{#1})}}    
\newcommand{\eqs}[1]{Eqs.~{\color{blue}(\ref{#1})}}  

\newcommand{\rsec}[1]{{\ref{#1}}}    
\newcommand{\secb}[1]{Sec.~{\ref{#1}}}
\newcommand{\secsb}[1]{Secs.~{\ref{#1}}}

    \makeatletter
    \renewcommand*{\@fnsymbol}[1]{\ensuremath{\ifcase#1
      \or {\color{DarkRed} *}
      \or {\color{DarkRed} **}
      \or {\color{DarkRed} ***}
      \or {\color{DarkRed}\normalsize \textcircled{\footnotesize a}}
      \or {\color{DarkRed}\normalsize \textcircled{\footnotesize b}}
      \or {\color{DarkRed}\normalsize \textcircled{\footnotesize c}}
      \or {\color{DarkRed}\normalsize \textcircled{\footnotesize d}}
      \or {\color{DarkRed}\normalsize \textcircled{\footnotesize e}}
      \or {\color{DarkRed}\normalsize \textcircled{\footnotesize f}}
      \or {\color{DarkRed}\normalsize \textcircled{\footnotesize g}}
      \or {\color{DarkRed}\normalsize \textcircled{\footnotesize h}}
      \or {\color{DarkRed}\normalsize \textcircled{\footnotesize i}}
      \or {\color{DarkRed}\normalsize \textcircled{\footnotesize j}}
      \or {\color{DarkRed}\normalsize \textcircled{\footnotesize k}}
      \or {\color{DarkRed}{3}}
      \or \ddagger
      \or \mathsection
      \or \mathparagraph
      \or \|
      \or **
      \or \dagger\dagger
      \or \ddagger\ddagger
      \else\@ctrerr\fi}}
    \makeatother

\begin{document}

\title{Statistical correlations in the oscillator model of quantum dissipative systems~}
\thanks{\small Revised version of the paper originally published as:\\
M. Patriarca,
{\it Statistical correlations in the oscillator model of quantum dissipative systems},
Il Nuovo Cimento B {\bf 111}, 61 (1996),
doi:~\href{http://link.springer.com/article/10.1007/BF02726201}{{\B {10.1007/BF02726201}}}.
}

\author{Marco Patriarca~}
\thanks{\small
  Email: {{\tt marco.patriarca}\,@\,kbfi.ee}\\
}
\affiliation{\href{www.kbfi.ee}{NICPB--National Institute of Chemical Physics and Biophysics, Tallinn, Estonia.}}



\date{\today}


\begin{abstract}
\noindent
{\bf Abstract.}
The problem of the initial conditions for the oscillator model of quantum dissipative systems is studied.
It is argued that, even in the classical case, the hypothesis that the environment is in thermal equilibrium implies a statistical correlation between environment oscillators and central system.
A simple form of initial conditions for the quantum problem, taking into account such a correlation in analogy with the classical ones, is derived on the base of symmetry considerations.
The same symmetries also determine unambiguously the form of the Lagrangian. 
As a check of the new form of correlated initial conditions (and of that of the Lagrangian), the problem of a forced Brownian particle under the action of arbitrary colored noise is studied: it is shown that one obtains an average position of a quantum wave packet equal to that of the corresponding classical Brownian particle.
Instead, starting from uncorrelated initial conditions based on the factorization hypothesis or from a different form of Lagrangian, non-physical results are obtained.
Similar considerations apply also to the mean square displacement. 
\end{abstract}


\maketitle


\section{Introduction}
The oscillator model of quantum dissipative systems~\cRef{Feynman1963a,Mazur1964a,Ullersma1966a, Ullersma1966b,Ullersma1966c,Ullersma1966d,Feynman1965a},
through the influence functional approach pioneered by Feynman and Vernon~\cRef{Caldeira1983a}, 
has been fruitfully applied to several problems, 
in which both quantum and statistical fluctuations play a significant role.
Important examples are provided by quantum Brownian motion~\cRef{Caldeira1983a,Schramm1987a,Grabert1988a}, 
dissipative tunneling~\cRef{Caldeira1983b,Leggett1987a},
and localization-delocalization transitions in periodic potentials~\cRef{Schmid1983a,Bulgadev1984a}.

In the oscillator model, the system under study, which will be assumed to have one degree of freedom $x$ and referred to as the \emph{central system}, is coupled to an infinite set of harmonic oscillators of coordinates $\bm{q} = \{ q_1, q_2, \ldots \}$, representing the environment.
The description of the central system at a generic time $t_b$ is made,
in the coordinate representation, through the reduced density matrix $\rho(x_b, x_b', t_b)$,
obtained by integrating the total density matrix $\rho(x_b, x_b', \bm{q}_b, \bm{q}_b', t_b)$
over the oscillator coordinates,
\begin{equation}
  \label{eq1}
  \rho(x_b, x_b', t_b) = \int d\bm{q}_b \, \rho(x_b, x_b', \bm{q}_b, \bm{q}_b, t_b) \, ,
\end{equation}
where $\int d\bm{q} \, (\dots) \equiv \prod_n \int dq_{n} \, (\dots)$.
In order to determine the time evolution law of the reduced density matrix,
one has to assign the initial conditions of the total system,
that is the total density matrix $\rho(x_a, x_a', \bm{q}_a, \bm{q}_a', t_a)$
at the initial time $t = t_a$~\cRef{Feynman1963a}.
A particular kind of initial conditions, considered previously in the literature~\cRef{Feynman1963a,Ullersma1966a, Ullersma1966b,Ullersma1966c,Ullersma1966d,Caldeira1983a},
is based on the factorization hypothesis,
\begin{equation}
  \label{eq2}
  \rho(x_a, x_a', \bm{q}_a, \bm{q}_a', t_a) = 
  \rho(x_a, x_a', t_a) \rho_\beta(\bm{q}_a, \bm{q}_a') \, .
\end{equation}
Here $\rho(x_a, x_a', t_a)$ is the density matrix of the central system and $\rho_\beta(\bm{q}_a, \bm{q}_a', t_a)$,
that represents the initial state of the environment,
describes oscillators in thermal equilibrium at an inverse temperature $\beta = 1 / k_\mathrm{B} T$.
According to \eq{eq2}, there is no statistical correlation between the central system and the environment
at $t = t_a$.
Unfortunately, the simple factorization hypothesis leads to nonphysical results, as shown in \secb{sec5}.

Later on, the importance of an initial correlation for the dynamics of the central system was recognized~\cRef{Hakim1985a}
and more general forms of correlated initial conditions were studied for the quantum problem~\cRef{Schramm1987a,Grabert1988a,Schramm1987a,MoraisSmith1990a,MoraisSmith1987a}.

As discussed below, the form of the initial conditions is closely connected to that of the total Lagrangian and to the initial state of the central system.
Studying the problem of the initial correlations between environment and central system implies studying also their coupling and therefore the corresponding Lagrangian form.
Starting from a total Lagrangian consistent with the Langevin equation, in this paper a new form of initial conditions is derived, in the simplifying hypothesis that the initial state of the central system is known --- also the case in which the initial quantum state of the central system is affected by some uncertainty is discussed.
The initial conditions thus obtained take into account the statistical correlation between environment and central system, following from the hypothesis that the environment is in thermal equilibrium, while keeping the mathematical simplicity of the uncorrelated conditions in \eq{eq2}.

In ~\secb{sec2} the classical model is summarized, discussing the form of the Lagrangian in relation to the Langevin equation and how even classically an initial correlation follows from the thermal equilibrium the environment.
In ~\secb{sec3} a novel form of initial conditions for the quantum problem is derived, partially based on the analogy with the classical model.
In \secb{sec4} the initial conditions obtained are applied to the study of the forced Brownian particle with arbitrary colored noise.
It is shown that the average position of a quantum wave packet coincides with that of the corresponding classical Brownian particle and that in the classical limit the quantum mean square displacement reduces to its classical counterpart.
In \secb{sec5} a detailed comparison between correlated and uncorrelated initial conditions is carried out in the particular case of white noise, showing that, starting from the uncorrelated initial conditions given by \eq{eq2}, one obtains nonphysical results both for the average motion and the spreading process of the wave packet.
The problem discussed and the solution suggested are summarized in terms of the underlying symmetries of the problem in \secb{conclusion}.

\section{Classical model}
\label{sec2}

In this section the classical oscillator model of linear dissipative systems is summarized.
The reason to start from the classical model is that
it provides a precious starting point for finding the initial conditions for the quantum model,
discussed below in \secb{sec3}.
The Lagrangian of the total system, composed by the central particle and the environment, is
\begin{eqnarray}
  \label{eq2_1}
  L(x, \dot{x}, \bm{q}, \dot{\bm{q}}, t) &=& 
  \frac{M}{2} \dot{x}^2 - V(x,t) 
  \nonumber \\
  &+& \sum_n \left\{ \frac{1}{2} m \dot{q}_n^2 - \frac{1}{2} m \omega_n^2 \left(q_n-x\right)^2 \right\},~~
\end{eqnarray}
where $\bm{q} = \{q_n\}$ are the coordinates of the oscillators, 
$\dot{\bm{q}} = \{\dot{q}_n\}$ their velocities,
$\omega_n/2\pi$ their frequencies, and $m$ their mass.
The constant $M$ and the function $V(x,t)$ represent the mass and the potential, respectively,
of the central degree of freedom $x$.
The effective equation of motion of $x$ can be obtained by eliminating the oscillator coordinates
from Lagrange's equations~\cRef{Zwanzig1973a} and turns out to be the generalized Langevin equation~\cRef{Mori1965a,Kubo1966a,Kubo1957a,Kubo1957b},
\begin{eqnarray}
  \label{eq2_2}
  M \ddot{x} + V'(x,t) = - \int_{t_0}^t ds \, K(t-s) \dot{x}(s) + R(t) \, .
\end{eqnarray}
Here $ V'(x,t) = \partial_x V(x,t)$,
while the memory kernel and the random force are given by
\begin{eqnarray}
  \label{eq2_3}
  K(\tau) &=& \sum_n m \omega_n^2 \cos \left(\omega_n \tau \right) \, ,
  \\
  \label{eq2_4}
  R(t) &=& \sum_n m \omega_n^2 \left\{ 
  \vphantom{\frac{\dot{q}}{\omega}}
  (q_{n0} - x_0) \cos \left[\omega_n (t - t_0) \right] \right. 
  \nonumber \\
  &&~~~~~~~~~~~~~~ 
  + \left. \frac{(\dot{q}_{n0} - \dot{x}_0)}{\omega_n} \sin \left[\omega_n (t - t_0) \right] \right\},~~
\end{eqnarray}
where $x_0 = x(t_0)$, $q_{n0} = q_n(t_0)$, and $\dot{q}_{n0} = \dot{q}_n(t_0)$.

The expression \req{eq2_4} for the random force $R(t)$ can be considered as a Fourier expansion
of a function with a zero-frequency average component equal to zero,
in which the oscillator initial coordinates and velocities determine the Fourier coefficients.
The random character of $R(t)$ requires $q_{n0}$ and $\dot{q}_{n0}$ to be random variables.
Therefore, statistical averages over the possible configurations of the stochastic force $R(t)$
are carried out as ensemble averages over the oscillator initial coordinates and velocities.
A most relevant property of the oscillator model is that, 
if the oscillators are assumed to be in thermal equilibrium at $t = t_0$, then
the classical fluctuation-dissipation theorem is automatically fulfilled
for any set of frequencies~\cRef{Zwanzig1973a} (it is in fact fulfilled
for each oscillator separately), i.e.,
\begin{eqnarray}
  \label{eq2_6}
  &&\langle R(t) \rangle = 0 \, ,
  \\
  \label{eq2_5}
  &&\beta \langle R(t) R(s) \rangle = K(t-s) \, ,
\end{eqnarray}
where $K(\tau)$ is given by \eq{eq2_3}.
Here the symbol $\langle \ldots \rangle$ represents a statistical average over the random force configurations,
i.e., over the initial conditions of the oscillators,
\begin{eqnarray}
  \label{eq2_5b}
  \langle \varphi \, \{q_{n0},\dot{q}_{n0}\} \rangle = 
  \int d\bm{q}_0  \, d\bm{\dot{q}}_0  \, \varphi\{q_{n0},\dot{q}_{n0}\} \, ,
\end{eqnarray}
where $\varphi\{q_{n0},\dot{q}_{n0}\}$ is a generic function of all the $\{q_{n0},\dot{q}_{n0}\}$
and $\int d\bm{q}_0  \, d\bm{\dot{q}}_0 = \prod_n \int dq_{n0} \int d\dot{q}_{n0}$;
the oscillator coordinates and velocities
are assumed to be distributed according to independent canonical probability densities,
\begin{eqnarray}
  \label{eq2_7}
  P_\beta(\bm{q}_0, \dot{\bm{q}}_0) = \prod_n P_\beta^{(n)}(q_{n0}-x_0, \dot{q}_{n0}) \, ,
\end{eqnarray}
where the function $P_\beta^{(n)}(q-x, \dot{q})$  is the canonical distribution of an oscillator
with frequency $\omega_n/2\pi$ and equilibrium position $q = x$,
\begin{equation}
  \label{eq2_8}
  P_\beta^{(n)}\!(q \!-\! x, \dot{q}) = \frac{\beta m \omega_n}{2 \pi} 
  \exp\left\{ \!
  -\frac{m\beta}{2} \! \left[ \dot{q}^2 + \omega_n^2 \left(q \!-\! x\right)^2 \right] 
  \! \right\} . ~
\end{equation}
The corresponding averages values are $\langle q \rangle = x$, $\langle \dot{q} \rangle = 0$,
while the variances are $m \omega_n^2 \langle (q-x)^2 \rangle = m \langle \dot{q}^2 \rangle = \beta^{-1}$.

The functions $P_\beta^{(n)}(q_{n0}-x_0, \dot{q}_{n0})$ depend parametrically 
on the initial central particle coordinate $x_0$;
this choice of the oscillator equilibrium positions naturally follows
from the form of the Lagrangian \req{eq2_1}~\cRef{Mazur1964a,Zwanzig1973a}.
It is to be noticed that \eq{eq2_8} implies a correlation 
$\langle q \, x \rangle \ne 0$ between the central system and the environment, 
since the probability density $P_\beta^{(n)}(q-x, \dot{q})$ 
cannot be factorized into the product of a function of $x$ times a function of $q$.

\begin{figure*}[ht]
 \centering
 \includegraphics[width=7cm]{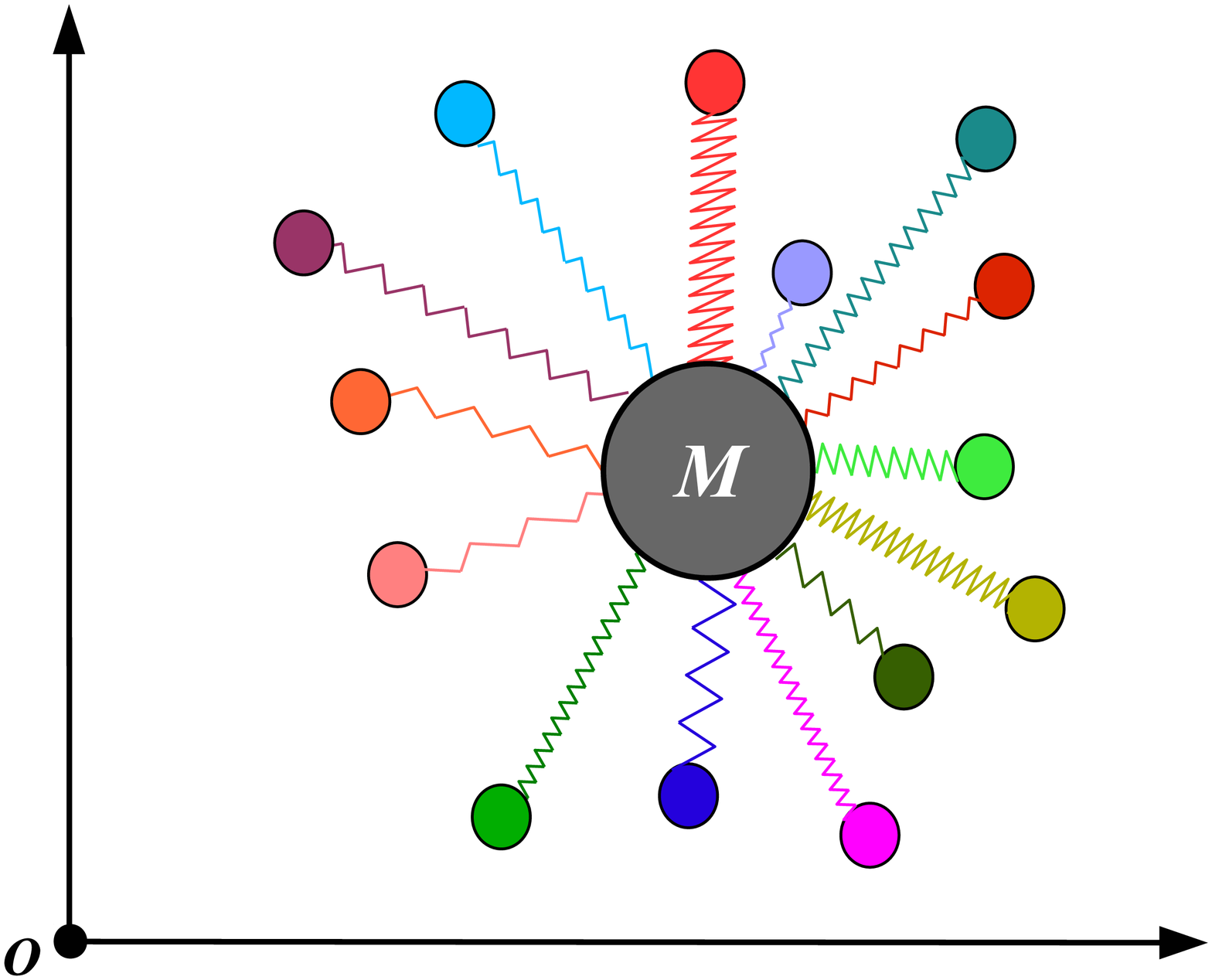}
 \includegraphics[width=7cm]{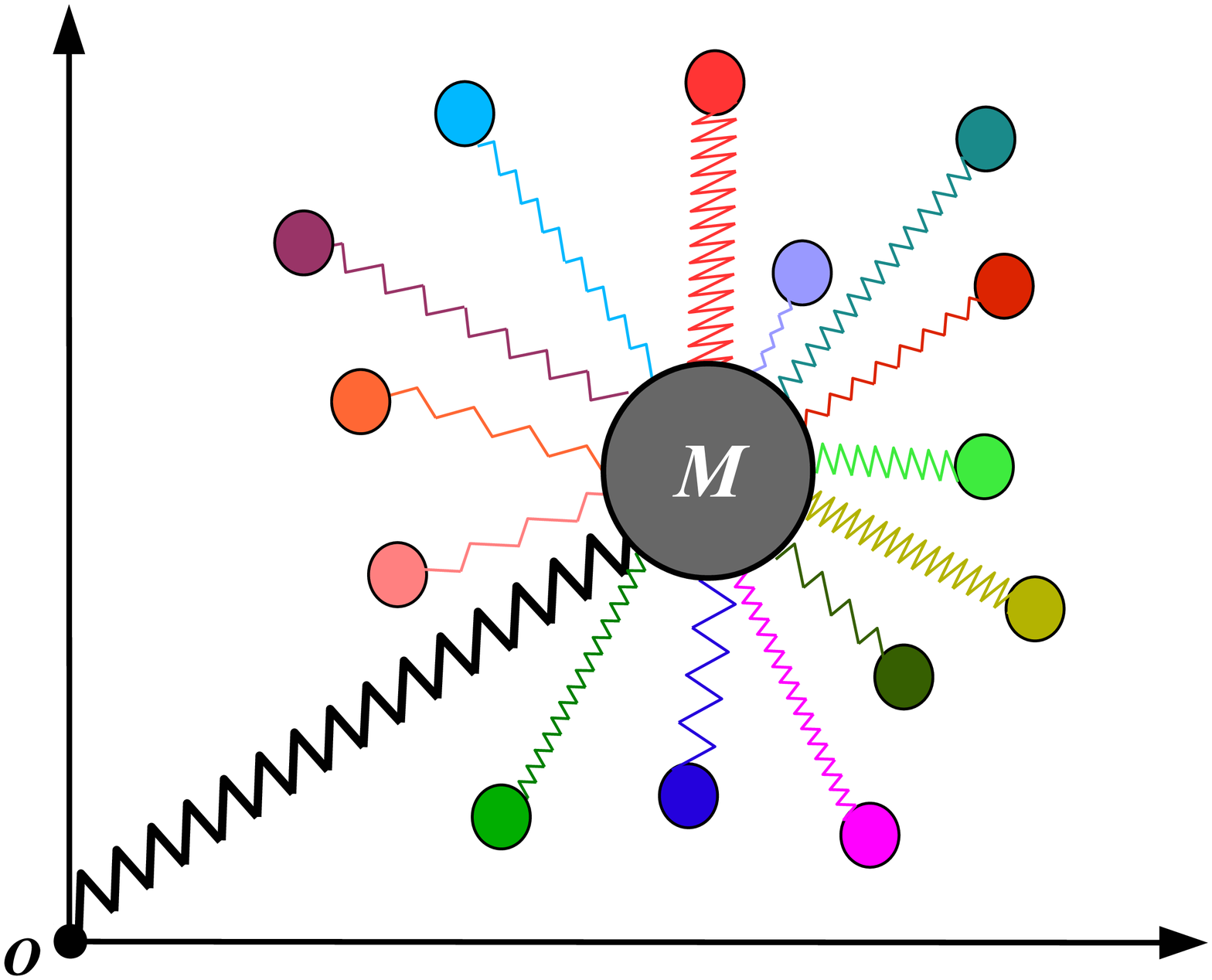}
 \caption{
 \label{scheme}
 Scheme of the different oscillator models corresponding to the Lagrangian in Eq.~(\ref{eq2_1}) (left) and Eq.~(\ref{eq2_1bis}) for $k_n = m \omega_n^2$ (right).
 In the latter case, the Lagrangian introduces a spurious attraction toward the origin of the coordinate system.
}
\end{figure*}

Dissipation can arise from the integral term in \eq{eq2_3}, 
when frequencies are continuously distributed,
so that the sums e.g. in \eq{eq2_3} over the discrete frequencies $\omega_n$ are replaced by the integral
\begin{eqnarray}
  \label{eq2_3_0}
   \sum_n \varphi_n 
   \to 
   \int_0^{+\infty} \!\!\! d\omega \, G(\omega) \varphi(\omega) \, ,
\end{eqnarray}
where $\varphi$ is a generic function of the angular frequency.
Here and in the following it is convenient anyway 
to refer to discrete sets of frequencies, taking
the continuous limit in the final results.
It is easy to check that the frequency distribution $G(\omega)$
is simply related to the power spectrum of the random force, as
\begin{eqnarray}
  \label{eq2_3_1}
  \mathcal{P}(\omega) 
  \equiv \int_0^{+\infty} d\tau \langle R(t) R(t+\tau) \rangle  
  = \frac{G(\omega)}{m\omega^2} \, .
\end{eqnarray}
As a relevant example, white noise is described by the frequency distribution function 
\begin{eqnarray}
  \label{eq5_1}
  G_\mathrm{W}(\omega) =  \frac{2 M \gamma}{\pi m} \frac{1}{\omega^2} \, ,
\end{eqnarray}
where $\gamma$ is the friction coefficient.
Correspondingly, from \eq{eq2_3} the correlation function is
\begin{eqnarray}
  \label{eq5_2}
  K_\mathrm{W}(\tau) 
  = \!\! \int_0^{+\infty} \!\!\! d\omega \, G_\mathrm{W}(\omega) \, m \omega^2 \cos(\omega\tau) 
  = 2 M \gamma \delta(\tau)  ,~~~~~
\end{eqnarray}
while the generalized Langevin equation \req{eq2_2} reduces to the Langevin equation
\begin{eqnarray}
  \label{eq5_3}
  M \ddot{x}(t) + V'(x,t) = - M \gamma \dot{x}(t) + R(t) \, ,
\end{eqnarray}
with $\langle R(t) \rangle = 0$ and $\langle R(t) R(s) \rangle = (2 M \gamma/\beta) \delta(t-s)$.
As another example, a general Gaussian noise has an exponential correlation function
\begin{eqnarray}
  \label{eq2_3c}
  K_\mathrm{L}(\tau) &=& 2 M \gamma \tau_\mathrm{L}^{-1} \exp(- \tau / \tau^{\vphantom{-1}}_\mathrm{L}) \, ,
\end{eqnarray}
which can be derived from the Lorentzian frequency distribution
\begin{eqnarray}
  \label{eq5_1b}
  G_\mathrm{L}(\omega) 
  =  \frac{2 M \gamma}{\pi m \, \omega^2} 
     \frac{1}{1 + (\omega \tau^{\vphantom{-1}}_\mathrm{L})^2} \, .
\end{eqnarray}
If the cutoff $\tau^{\vphantom{-1}}_\mathrm{L}$ is much smaller than any natural time scale of the system,
one can neglect $1/[1 + (\omega \, \tau^{\vphantom{-1}}_\mathrm{L})^2]$ and the white noise limit is recovered.

The physical picture emerging from the total Lagrangian \req{eq2_1}
and the initial conditions given by \eqs{eq2_7}-\req{eq2_8}
underlies the so-called oscillator model of dissipative systems.
In the mechanical interpretation of the model, the Brownian particle is represented as a bare central particle coupled to an infinite set of oscillators~\cRef{Schramm1987a,Grabert1988a}:
the average value of the total force produced by the oscillators causes a dissipative force, while their oscillations perturb the central particle trajectory in an erratic way and account for the environmental noise.

The correspondence and internal consistency between the Lagrangian (\ref{eq2_1}) and the initial conditions \eqs{eq2_7}-\req{eq2_8} is to be noticed, in the sense that the second ones follow from the first one under the hypothesis of thermal equilibrium of the environment oscillators.
Different forms of Lagrangian,  such as the commonly used one,
\begin{eqnarray}
  \label{eq2_1bis}
  L(x, \dot{x}, \bm{q}, \dot{\bm{q}}, t) &=& 
  \frac{M}{2} \dot{x}^2 - V(x,t) 
  \nonumber \\
  &+& \sum_n \left\{ \frac{m}{2}  \dot{q}_n^2 - \frac{m \omega_n^2}{2} q_n^2 + k_n q_n x \right\},~~
\end{eqnarray}
where the $k_n$'s are coupling constants~\cite{Ullersma1966a, Ullersma1966b,Ullersma1966c,Ullersma1966d,Caldeira1983a}, represent peculiar systems and provide in general wrong predictions.
For instance, in the particular case $k_n = m \omega_n^2$, it is easy to show, by rearranging the various terms, that according to Lagrangian (\ref{eq2_1bis}) the central particle is harmonically bound to the origin, with a coupling constant $\sum_n m \omega_n^2 $, so that the origin would be assigned a special role.
The two mechanical models, corresponding to the Lagrangians (\ref{eq2_1}) and (\ref{eq2_1bis}) with $k_n = m \omega_n^2$, are compared in Fig.~\ref{scheme}.

\section{Quantum model}
\label{sec3}

The study of the classical model gives us useful information to state the initial conditions for the quantum problem.
In particular, it suggests that the hypothesis of thermal equilibrium has to be made \emph{only}
for the environment oscillators $\bm{q}$ and one has to beware of the fact that their equilibrium position
is given by $x$, the coordinate of the central system.
As long as the central system is concerned, it will have, in general, arbitrary initial conditions.
Thus, the initial density matrix at $t = t_a$ is
\begin{equation}
  \label{eq3_1}
  \rho(x_a, x_a', \bm{q}_a, \bm{q}_a', t_a) = 
  \rho(x_a, x_a', t_a) \rho_\beta(\bm{q}_a, \bm{q}_a'; x_a, x_a') \, .
\end{equation}
Here $\bm{q}_a = \{ q_{na} \}$ are the oscillator coordinates,
$\rho(x_a, x_a', t_a)$ represents the (arbitrary) initial state of the central system,
and $\rho_\beta(\bm{q}_a, \bm{q}_a'; x_a, x_a')$, that depends parametrically 
also on the variables $x_a$ and $x_a'$, is the initial density matrix of the environment 
in thermal equilibrium.
On the analogy of the classical initial conditions defined by \eq{eq2_6}, here it is assumed that
\begin{eqnarray}
  \label{eq3_2}
  \rho_\beta(\bm{q}_a, \bm{q}_a'; x_a, x_a') 
  = 
  \prod_n \rho_\beta^{(n)}(q_{na} - X_a, q_{na}' - X_a) \, ,
\end{eqnarray}
where $X_a$ represents the coordinate of the central system at $t = t_a$
and $\rho_\beta^{(n)}(q - X, q' - X)$ is the equilibrium density matrix of an oscillator
with frequency $\omega_n/2\pi$ and equilibrium position $q = X$.
The explicit expression of $\rho_\beta^{(n)}(q - X, q' - X)$ can be found from that of
$\rho_\beta^{(n)}(q, q')$, the equilibrium density matrix of an oscillator with equilibrium position $q = 0$~\cRef{Feynman1965a},
by carrying out the translation transformation $q \to q - X$, $q' \to q' - X$,
\begin{eqnarray}
  \label{eq3_3}
  &&\!\!\!\!\!
  \rho_\beta^{(n)}(q - X, q - X)   = \mathcal{F}_n(\beta)
  \nonumber \\
  &&\!\!\!\!\!
  \times \! \exp \! \left\{ \!
    -  \frac{m\omega_n}{2\hbar}
    \! \left[
    \! \frac{(q \!-\! X)^2 \!+\! (q' \!-\! X)^2}{\tanh(\beta\hbar\omega_n)}
    \!-\! 
    \frac{2 (q \!-\! X)(q' \!-\! X)}{\sinh(\beta\hbar\omega_n)}
    \! \right]
  \! \right\},
  \nonumber \\
\end{eqnarray}
where $\mathcal{F}_n(\beta) = \sqrt{m\omega_n / \pi\hbar\coth(\beta\hbar\omega_n/2)}$ is a normalization factor
such that $\int dq \, \rho_\beta^{(n)}(q, q)=1$.
The coordinates of the central system at $t = t_a$, represented by $X_a$ in \eq{eq3_2}, 
can only depend on $x_a$ and $x_a'$.
The problem of determining the initial conditions for the quantum problem
is thus reduced to that of finding $X_a$ in terms of $x_a$ and $x_a'$.
For reasons of translation and reflection invariance, $X_a = c x_a + (1 - c) x_a'$, where $c$ is a constant.
Since the $\rho_\beta^{(n)}(q - X, q - X)$'s represent equilibrium states, 
they must be left unchanged by a time-reversal operation, in which the variables $q$ and $x$ are interchanged
with the variables $q'$ and $x'$. 
It follows that $c = 1/2$, so that
\begin{eqnarray}
  \label{eq3_4}
  X_a = \frac{x_a + x_a'}{2} \, .
\end{eqnarray}
This expression can be obtained more easily from the definition of average coordinate
of a system described by a density matrix, as defined by Schmid~\cRef{Schmid1982a},
and will be checked in \secsb{sec4} and \rsec{sec5}.

Equations \req{eq3_1}-\req{eq3_4} imply an initial correlation between the central system and the environment,
because the dependence of the $\rho_\beta^{(n)}(q - X, q - X)$'s on the $q$ and $x$ variables 
cannot be factorized.
For this reason, even if the total density matrix at $t = t_a$ can be written as the product of the density matrices
of the subsystems, in the following the initial conditions defined by \eqs{eq3_1}-\req{eq3_4} will be referred to as \emph{correlated initial conditions}.

By using the influence functional approach~\cRef{Feynman1963a,Feynman1965a,Caldeira1983a}, with the initial conditions illustrated above,
one can show that the reduce density matrix $\rho(x,x',t)$ evolves with time according to the integral equation 
\begin{equation}
  \label{eq3_5}
  \rho(x_b, x_b', t_b) 
  \!=\! 
  \! \int \!\! d x_a \!\! \int \!\! d x_a' J(x_b, x_b', t_b|x_a, x_a', t_a) \rho(x_a, x_a', t_a),
\end{equation}
where $\rho(x_a, x_a', t_a)$ represents the (arbitrary) initial conditions of the central system
and the effective propagator can be written as
\begin{equation}
  \label{eq3_6}
  J(x_b, x_b', t_b|x_a, x_a', t_a) 
  = \int \mathcal{D}x \int \mathcal{D}x'\exp \left( \frac{i}{\hbar} A[x,x'] \right) \, .
\end{equation}
The effective action $A[x,x']$ is given by
\begin{equation}
  \label{eq3_7}
  A[x,x'] = S[x] - S[x'] + \hbar \, \Phi[x,x'] \, ,
\end{equation}
where $S[x]$ is the action of the isolated central system,
\begin{equation}
  \label{eq3_8}
  S[x] = 
  \int_{t_a}^{t_b} dt 
  \left[ 
  \frac{1}{2} M \dot{x}(t)^2 - V \left( x(t), t \vphantom{t^A_B} \right) 
  \right] \, ,
\end{equation}
and $\Phi[x,x']$ is the influence phase~\cRef{Feynman1963a,Caldeira1983a},
\begin{eqnarray}
  \label{eq3_9}
  &&\!\!\!\!\!
  \hbar \, \Phi[x,x']
  \nonumber \\
  && =
  \frac{1}{2} \int_{t_a}^{t_b} dt \int_{t_a}^t ds [x'(t) - x(t)] K(t-s) [\dot{x}(s) + \dot{x}'(s)]
  \nonumber \\
  && +
  \frac{i}{\hbar} \int_{t_a}^{t_b} dt \int_{t_a}^t ds [x'(t) - x(t)] \alpha(t-s) [x'(s) - x(s)],
  \nonumber \\
\end{eqnarray}
which represents the interaction with the environment.
Here $K(\tau)$ is the correlation function defined by \eq{eq2_3} and $\alpha(\tau)$ is given by the expression
\begin{eqnarray}
  \label{eq3_10}
  \alpha(\tau) = 
  \sum_n \frac{1}{2} m \hbar \omega_n^3 \coth \left(\frac{\beta \hbar \omega_n}{2}\right) \cos(\omega_n \tau) \, ,
\end{eqnarray}
which reduces to $\beta^{-1} K(\tau)$ in the large-temperature limit.

The considerations made above and, in particular, \eqs{eq3_4} and \req{eq3_9}
are valid for homogeneous dissipative systems with additive noise,
but they can be generalized to the case of inhomogeneous dissipative systems with multiplicative noise~\cRef{Illuminati1994a}.

\section{Forced {B}rownian particle}
\label{sec4}
%

In this section the dynamics of a Brownian particle, 
under the action of an arbitrary external bias $f(t)$ and 
a colored noise $R(t)$ with correlation function $K(\tau$), is studied;
the external potential is then $V \left( x(t), t \vphantom{t^A_B} \right) \equiv - x f(t)$.

\subsection{Classical problem}
\label{sec4a}
The generalized Langevin equation for a forced Brownian particle reads
\begin{eqnarray}
  \label{eq4_1}
  M \ddot{x}(t) = - \int_{t_0}^t ds \, K(t-s) \, \dot{x}(s) + f(t) + R(t) \, ,
\end{eqnarray}
where $f(t)$ represents the external force.
It is useful to introduce the kernel (Green function) $I(\tau)$  of the equation,
defined as the solution of the associated homogeneous model [for $f(t) \equiv 0$], i.e., 
\begin{eqnarray}
  \label{eq4_2}
  M \frac{\partial I(\tau)}{\partial \tau} + \int_{0}^\tau d\sigma \, I(\tau - \sigma) K(\sigma) = 0 \, .
\end{eqnarray}
Multiplying both sides of \eq{eq4_1} by $I(u - t)$,
integrating between $t = t_0$ and $t = u$, and using \eq{eq4_2} one obtains
(after a renaming of the variables)
\begin{eqnarray}
  \dot{x}(t) = v_0 \, I(t - t_0) \!&+&\! \frac{1}{M} \int_{t_0}^t \!\! ds \, I(t - s) f(s) ,
  \nonumber 
  \\
             \!&+&\! \frac{1}{M} \int_{t_0}^t ds \, I(t - s) R(s) \, ,
  \label{eq4_3}
\end{eqnarray}
where $v_0 = \dot{x}(t_0)$ is the initial velocity and the normalization
\begin{eqnarray}
  \label{eq4_4}
  I(0) = 1 \, ,
\end{eqnarray}
for $I(\tau)$ has been assumed.
Integrating once more in time \eq{eq4_3} provides the solution for the central particle coordinate,
\begin{eqnarray}
  x(t) = x_0 \!&+&\! v_0 \, \Delta(t - t_0) + \frac{1}{M} \int_{t_0}^t ds \, \Delta(t - s) f(s) 
  \nonumber \\
             \!&+&\! \frac{1}{M} \int_{t_0}^t ds \, \Delta(t - s) R(s) \, ,
  \label{eq4_5}
\end{eqnarray}
where $x_0 = x(t_0)$ and $\Delta(\tau)$, the integral function  of $I(\tau)$, was introduced,
\begin{eqnarray}
  \label{eq4_6}
  \Delta(\tau) = \int_{0}^\tau d\sigma \, I(\sigma) \, .
\end{eqnarray}
For consistence with the definitions of $x_0$ and $\dot{x}_0$, the initial conditions for $\Delta(\tau)$ are
\begin{eqnarray}
  \label{eq4_7}
  \Delta(0) = 0 \, ,
  \\
  \label{eq4_8}
  \dot{\Delta}(0) = 1 \, ,
\end{eqnarray}
where $\dot{\Delta}(\tau) = \partial\Delta(\tau)/\partial\tau$.
It is straightforward to check that in the limit of small dissipation $K(\tau) \approx 0$, 
when the interaction with the environment is negligible,
from \eq{eq4_2} and \req{eq4_4} one obtains that $I(\tau) \equiv 1$;
then, from \eq{eq4_6}, one has $\Delta(\tau) = \tau$.
Correspondingly, the velocity and position given by \eqs{eq4_3} and \req{eq4_5}
reduce to the solution of a forced particle with no dissipation.

Using $\langle R(t) \rangle = 0$,
the average value of the velocity $\langle\dot{x}(t)\rangle$ and
of the coordinate $\langle x(t)\rangle$ is given by the first three terms on the right hand side
of \eqs{eq4_3} and \req{eq4_5}, respectively.
 
If the initial state of the central particle is affected by some statistical uncertainties, 
described by a probability density $P^{(0)}(x_0, v_0)$
for the initial position $x_0$ and velocity $v_0$,
this can be taken into account in the calculation of the average values
by performing an additional average.
As a simple example and in view of a comparison with the quantum case,
here its is assumed that the central particle position $x_0$ is known with some uncertainty
and is a random variable distributed according to
\begin{eqnarray}
  P^{(0)}(x_0) = 
  \frac{1}{\sqrt{2\pi}\sigma_0} 
  \exp \left[\! - \frac{(x_0 - \bar{x}_0)^2}{2\sigma_0^2} \! \right] \! ,
  \label{eq4_8b}
\end{eqnarray}
with average value $\bar{x}_0$ and variance $\sigma_0^2$.
The only effect in the average position $\langle x(t)\rangle$ 
of the additional average over $x_0$ is to replace the initial value $x_0$ 
with its average value , i.e.,
\begin{equation}
  \langle x(t) \rangle = \bar{x}_0 + v_0 \, \Delta(t - t_0) + \frac{1}{M} \int_{t_0}^t ds \, \Delta(t - s) f(s) \, ,
  \label{eq4_8c}
\end{equation}
where the symbol $\langle\ldots\rangle$ represents a statistical average both on the random force
and on the initial coordinate $x_0$.
In the mean square displacement $\langle \delta x(t)^2 \rangle$, 
the uncertainty on the initial coordinate produces an additional initial contribution equal to $\sigma_0^2$.
Performing both averages, one obtains
\begin{eqnarray}
  &&\langle \delta x(t)^2 \rangle 
  \equiv \langle [x(t) - \langle x(t) \rangle]^2 \rangle
  \nonumber
  \\
  &&= \sigma_0^2
  + \frac{1}{M\beta} \int_{t_0}^t \!\! ds \!\! \int_{t_0}^{t} \!\! du \, \Delta(s) K(|s-u|) \Delta(u) .
  \label{eq4_8d}
\end{eqnarray}
%

\subsection{Quantum problem}
Here the quantum problem is considered.
If the quantum state of the central particle at the initial time $t = t_a$ is known
the density matrix can be written as
\begin{eqnarray}
  \label{eq4_9}
  \rho(x_a, x_a', t_a) = \psi( x_a, t_a) \psi^*( x_a', t_a) \, ,
\end{eqnarray}
where $\psi( x_a, t_a)$ is the initial wave function.
For simplicity a Gaussian wave packet is assumed,
\begin{equation}
  \label{eq4_10}
  \psi( x_a, t_a) 
  \!=\! 
  \frac{1}{(2\pi\sigma_0^2)^{1/4}} 
  \exp \left[\! - \frac{(x_a - x_0)^2}{4\sigma_0^2} \!+\! i\frac{M v_0}{\hbar} x_a \! \right] \! ,
\end{equation}
with $\int dx_a |\psi(x_a,t_a)|^2 = 1$.
Here the parameters $x_0$ and $v_0$ represent the initial average position and velocity, respectively,
\begin{eqnarray}
  \label{eq4_10b}
  \!\!\!\!\!\!\!\!\!\!\!\!\!\!\!\!&&\langle x \rangle = \int dx_a |\psi(x_a,t_a)|^2 = x_0 \, 
  \\ 
  \!\!\!\!\!\!\!\!\!\!\!\!\!\!\!\!&&\langle \hat{v} \rangle \! \equiv \! \left\langle \! \frac{\hat{p}}{M} \! \right\rangle 
  \! = \!\!\! \int \!\! dx_a \psi(x_a,t_a)
  \!\left( \! -\frac{i\hbar}{M} \frac{\partial}{\partial x_a} \! \right) \! \psi(x_a,t_a) = v_0,
  \nonumber \\
\end{eqnarray}
while the parameter $\sigma_0$ defines the corresponding uncertainties, 
\begin{eqnarray}
  \label{eq4_10c}
  &&\langle \delta x^2 \rangle 
  \equiv 
  \left\langle [x_a - \langle x \rangle ]^2 \right\rangle 
  = \sigma_0^2 \, ,
  \\
  &&\left\langle \delta \hat{p}^2 \right\rangle 
     \equiv 
     \left\langle [\, \hat{p} - \langle \hat{p} \rangle ]^2 \right\rangle 
     = 
     \hbar^2 / 4 \sigma_0^2 \, .
\end{eqnarray}
For the study of the quantum problem it is convenient to introduce the coordinates
\begin{eqnarray}
  \label{eq4_10d}
  X(t) &=& \frac{x(t) + x'(t)}{2} \, ,
  \\
  \label{eq4_10e}
  \xi(t) &=& x'(t) - x(t) \, .
\end{eqnarray}
The physical meaning of $X$ and $\xi$ appears clearly below.
The density matrix expressed in the new variables, from \req{eq4_9} and \req{eq4_10}, is
\begin{eqnarray}
  \label{eq4_9b}
  &&\rho(X_a, \xi_a, t_a) = 
  \nonumber
  \\
  &&\frac{1}{ \sqrt{2\pi}\sigma_0 } 
    \exp \left[\! 
       - \frac{(X_a - x_0)^2}{2\sigma_0^2} 
       - \frac{\xi_a^2}{8\sigma_0^2} 
       \!-\! i\frac{M v_0}{\hbar} \xi_a \! 
    \right] \! .
\end{eqnarray}
The effective propagator given by \eq{eq3_6} now becomes
\begin{equation}
  \label{eq3_6b}
  J(X_b, \xi_b, t_b|X_a, \xi_a, t_a) 
  = \int \mathcal{D}X \int \mathcal{D}\xi \exp \left( \frac{i}{\hbar} A[X,\xi] \right) \, ,
\end{equation}
where the effective action, from \eqs{eq3_7}-\req{eq3_9}, is
\begin{eqnarray}
  \label{eq4_11}
  &&A[X,\xi]
  = \int_{t_a}^{t_b} \!\! dt 
  \left[
     - M \dot{X}(t) \dot{\xi}(t) - f(t)\xi(t) \vphantom{\frac{i}{\hbar}}
  \right]
  \nonumber \\
  &&+ \int_{t_a}^{t_b} \!\!\! dt \int_{t_a}^t \! ds \, \xi(t) 
  \left[ 
     K(t-s) \dot{X}(s) + \frac{i}{\hbar} \, \alpha(t-s) \xi(s) 
  \right] \! . ~~~~~~~~
\end{eqnarray}
Since this functional is quadratic in $X(t)$ and $\xi(t)$, 
the effective propagator can be evaluated as~\cRef{Feynman1963a,Feynman1965a}
\begin{eqnarray}
  \label{eq4_12}
  &&J(X_b, \xi_b, t_b|X_a, \xi_a, t_a) 
  \nonumber
  \\
  &&= \mathcal{F}(\tau) \exp \left[ \frac{i}{\hbar} A_\mathrm{cl}(X_b, \xi_b, t_b|X_a, \xi_a, t_a) \right] \, .
\end{eqnarray}
Here $\tau = t_b - t_a$, $\mathcal{F}(\tau)$ is a normalization factor, 
and $A_\mathrm{cl}(X_b, \xi_b, t_b|X_a, \xi_a, t_a)$ is
the effective action $A[X,\xi]$ computed along the classical trajectories $X(t)$ and $\xi(t)$
defined by $\delta A[X,\xi]/\delta X(t) = 0$ and $\delta A[X,\xi]/\delta \xi(t) = 0$,
i.e., from \eqs{eq3_7}-\req{eq3_9},
\begin{eqnarray}
  M \ddot{X}(t) \!&=&\! 
  - \! \int_{t_a}^t \!\!\! ds \, \!\!\left[\!
      K(t \!-\! s) \dot{X}(s) + \frac{i}{\hbar} \alpha(|t \!-\! s|) \xi(s)
  \!\right]\!
  + f(t) \, ,
  \nonumber \\
  \label{eq4_13}
  \\
  \label{eq4_14}
  M \ddot{\xi}(t) &=& 
  \int_{t_a}^t ds \, \dot{\xi}(s) K(t-s) 
  - \xi_b K(t_b - t) \, ,
\end{eqnarray}
with boundary conditions $X(t_a) = X_a$, $\xi(t_a) = \xi_a$, $X(t_b) = X_b$, and $\xi(t_b) = \xi_b$.
It is to be noticed that both \eqs{eq4_13} and \req{eq4_14} have the same mathematical structure 
of \eq{eq4_1} and can be solved in a similar way.

Integrating by parts in the variable $X$ the first term in the integral in \eq{eq4_11}
and using the first classical equation \req{eq4_13}, the effective action can be simplified as
\begin{eqnarray}
  \label{eq4_15}
  A[X,\xi] &=& - M [ \dot{X}_b \xi_b - \dot{X}_a \xi_a ] 
  \nonumber \\
  && - \frac{i}{2\hbar} \int_{t_a}^{t_b} dt \int_{t_a}^t ds \, \xi(t) \, \alpha(t-s) \xi(s) \, ,
\end{eqnarray}
where $\dot{X}_a = X(t_a)$ and $\dot{X}_b = X(t_b)$.
A further simplification is possible if one is interested in the probability density 
$P(X_b, t_b) = \rho(X_b, \xi_b\!=\!0, t_b)$, since in this case the solutions
of the classical equations with boundary condition $\xi_b = 0$ are needed. 

From the solution of the classical equation \req{eq4_14} for $\xi(t)$ one obtains
\begin{eqnarray}
  \label{eq4_16a}
  &&\left. \vphantom{\dot{\xi}_b} \xi(t) \right|_{\xi_b = 0} = - \dot{\xi}_b \, \Delta(t-t_a) \, ,
  \nonumber \\
  \label{eq4_16b}
  &&\left. \dot{\xi}_b \right|_{\xi_b = 0} = - \dot{\xi}_a \, \Delta(\tau) \, .  
\end{eqnarray}
By replacement in the solution of \eq{eq4_13} one obtains
\begin{eqnarray}
  \!\!\!\!\!\!\!\!&&\left. X(t) \right|_{\xi_b = 0} = 
  X_a + \dot{X}_a \, \Delta(t-t_a) 
  + \frac{1}{M} \!\! \int_{t_a}^t ds \, \Delta(t-s) f(s)
  \nonumber \\
  \label{eq4_17}
  \!\!\!\!\!\!\!\!&& - \frac{i \xi_a}{M \hbar \, \Delta(\tau)} 
  \! \int_{t_a}^t \!\! ds \!\! \int_{t_a}^{t_b} \!\! du \, \Delta(t-s) \alpha(|s-u|) \Delta(t_b-u) .~~
\end{eqnarray}
Computing this expression at $t = t_b$ and inverting, one obtains
\begin{eqnarray}
  \label{eq4_18}
  \left. \dot{X}_a \right|_{\xi_b = 0} = 
  \frac{X_b-X_a}{\Delta(\tau)} 
  + \frac{i \epsilon(\tau) \xi_a }{\hbar \, \Delta(\tau)^2}  
  - \frac{1}{M \, \Delta(\tau)} \eta[f]  , 
\end{eqnarray}
where the following functionals were defined,
\begin{eqnarray}
  \label{eq4_19}
  &&\epsilon(\tau) = \int_{0}^\tau \!\! ds \!\! \int_{0}^{\tau} \!\! du \, \Delta(s) \alpha(|s-u|) \Delta(u) ,
  \\
  \label{eq4_20}
  &&\eta[f] = \int_{t_a}^{t_b} ds \, \Delta(t_b-s) f(s) \, .
\end{eqnarray}
Then, the final expression for the effective propagator \req{eq4_12}
computed for $\xi_b=0$ is given by
\begin{eqnarray}
  J(X_b, 0, t_b &|& X_a, \xi_a, t_a) \!=\! \mathcal{F}(\tau) 
  \exp\left\{ 
  \frac{iM}{\hbar \, \Delta(\tau)} (X_b-X_a) \xi_a 
  \right.
  \nonumber \\
  \label{eq4_21}
  &\vphantom{.}&\left.
  - \frac{i}{2 \hbar \, \Delta(\tau)} \eta[f] \xi_a
  - \frac{M \epsilon(\tau)}{2 \hbar^2  \Delta(\tau)^2} \xi_a^2  
  \right\} \, .
\end{eqnarray}
Using \eq{eq3_5} and performing the integrations on the initial variables $X_a$ and $\xi_a$
one obtains a Gaussian form for the probability density at $t = t_b$,
\begin{eqnarray}
  &&P(X_b, t_b) = \rho(X_b, \xi_b\!=\!0, t_b) 
  \nonumber \\
  &&\!=\! 
  \! \int \!\! d X_a \!\! \int \!\! d \xi_a \, J(X_b, 0, t_b|X_a, \xi_a, t_a) \, \rho(X_a, x\xi_a, t_a)
  \nonumber \\
  &&= \mathcal{N(\tau}) \exp \left\{ 
      - \frac{\left[X_b - \langle X(t_b) \rangle \right]}{2\langle\delta X^2(\tau)\rangle} 
  \right\} \, .
  \label{eq4_22}
\end{eqnarray}
Here $\langle X(t_b) \rangle$ is the average position and is given by
\begin{eqnarray}
  \label{eq4_23}
  \langle X(t_b) \rangle = x_0 + v_0 \, \Delta(\tau) + \eta[f] \, .
\end{eqnarray}
It is to be noticed that the average motion of the wave packet is classical,
as expected for quadratic actions; in fact, this function represents 
the classical solution given by \eq{eq4_5};
the formal equivalence with \eq{eq4_5} is obtained by replacing
the proper initial and final times, i.e., $t_a$ with $t_0$ and $t_b$ with $t$,
and the  initial quantum average values of position and velocity 
$\langle X(t_0) \rangle$ and $\langle \dot{X}(t_0) \rangle$
with the corresponding classical initial values $x_0 $ and $v_0$.

The variance of the Gaussian distribution provides the mean square displacement,
\begin{eqnarray}
  \label{eq4_24}
  &&\langle\delta X^2(\tau)\rangle = \sigma^2_0
  + \left[ \frac{M \Delta(\tau) }{2 M \sigma_0} \right]^2
  + \frac{\epsilon(\tau)}{M}  \, .
\end{eqnarray}
The first term on the right hand side represents the initial quantum uncertainty on the particle position,
see \eqs{eq4_10} and \req{eq4_10c}.
Also the second term has a quantum origin.
Its form, however, through the function $\Delta(\tau)$, is strongly affected by the environment;
in the limit of small dissipation, in which $\Delta(\tau) \to \tau$,
one recovers the time-dependent part of the mean square displacement of a free quantum particle,
given by $(\hbar\tau/2M\sigma_0)^2$.
The last term on the right hand side of \eq{eq4_24} originates from by thermal fluctuations.
It can be checked that in the limit of large temperatures, in which $\alpha(\tau) \approx K(\tau)/\beta$, 
and using \eq{eq4_24}, it reduces to the mean square displacement 
of a classical Brownian particle, given by the second term on the right hand side of \eq{eq4_8d}.

Finally, $\mathcal{N}(\tau)$ is a normalization factor given by
\begin{eqnarray}
  \label{eq4_25}
  &&\mathcal{N}(\tau) = \frac{ \hbar \, \Delta(\tau) \mathcal{F}(\tau) }{M}
  \sqrt{ \frac{2\pi}{\sigma(t)^2} } \, .
\end{eqnarray}
Since normalization of the probability density \req{eq4_22}
requires that $\mathcal{N}(\tau) = 1/\sqrt{2\pi\langle\delta X^2(\tau)\rangle}$,
comparison with \eq{eq4_25} provides the normalization factor $\mathcal{F}(\tau)$ 
for the effective propagator \req{eq4_21}, 
\begin{eqnarray}
  \label{eq4_26}
  \mathcal{F}(\tau) = \frac{ M } { 2 \pi \hbar \, \Delta(\tau) } \, .
\end{eqnarray}
In the limit of small dissipation mentioned above, in which $\Delta(\tau) \to \tau$,
one recovers the normalization factor of the density matrix propagator of a free quantum particle,
$\mathcal{F}(\tau) \to M/2\pi\hbar\tau$.

\section{Uncorrelated versus correlated initial conditions}
\label{sec5}
%
The basic differences between uncorrelated and correlated initial conditions for the dynamics of a Brownian particle
are here illustrated for a specific example.
For simplicity, a free Brownian particle, in the limits of large temperature and white noise, is considered.
Notice that these two limits are distinct and interdependent.
The large temperature limit holds when $\beta^{-1} \gg \hbar\Omega$, where $\Omega$ the cutoff frequency
of the oscillator density $G(\omega)$.
In this limit the kernel $\alpha(\tau) \to \beta^{-1} K(\tau)$, see \eq{eq3_10}, i.e.,
one recovers the classical fluctuation-dissipation theorem.
The white noise limit is recovered when the noise correlation time $\approx \Omega^{-1}$
is much smaller than any relaxation time scale $\tau_\mathrm{rel}$ of the  system, i.e., 
$\Omega^{-1} \ll \tau_\mathrm{rel}$,
and the noise becomes $\delta$-correlated, $K(\tau) \propto \delta(\tau)$.
These two conditions have to hold at the same time,
$\tau_\mathrm{rel}^{-1} \ll \Omega \ll 1/(\beta\hbar)$.

As shown in the preceding section, the solution of the classical macroscopic equation,
$M \ddot{X}_\mathrm{av}(t) = - M \gamma \dot{X}_\mathrm{av}(t)$,
coincides with the average position $\langle x(t) \rangle$ of a Gaussian quantum wave packet
with initial average coordinate $\langle x(t_0) \rangle = x_0 = x_\mathrm{av}(t_0)$ 
and average velocity $\langle \hat{p}(t_0) \rangle/M = v_0 = \dot{x}_\mathrm{av}(t_0)$.
The macroscopic solution for this case is given by the general solution in \eq{eq4_5}, where
\begin{eqnarray}
  \label{eq5_4}
  \Delta(\tau) = - \frac{1}{\gamma} \left[1 - \exp(-\gamma\tau) \right] \, .
\end{eqnarray}
Its asymptotic limit is $\Delta(\tau \! \gg \! \gamma^{-1}) = 1/\gamma$,
so that the difference between final and initial position of the classical particle, 
as well of the center of mass of the quantum wave packet, is
\begin{eqnarray}
  \label{eq5_5}
          L = X_\mathrm{av}(+\infty) - X_\mathrm{av}(t_0) =  \frac{v_0}{\gamma} \, .
\end{eqnarray}
The modulus of $L$ provides the overall distance covered by the particle.

As for the mean square displacement of the quantum particle,
it is given by the variance of the Gaussian wave packet.
By solving \eqs{eq4_13} and \req{eq4_14} in the white noise and large temperature limit,
for the function $\epsilon(\tau)$ defined by \eq{eq4_19} one obtains 
\begin{equation}
  \label{eq5_6}
  \epsilon(\tau) = \frac{2}{\beta\gamma^2} 
  \! \left[ 
  \gamma\tau - \frac{3}{2} + 2 \exp(-\gamma\tau) - \frac{1}{2}  \exp(-2\gamma\tau)
  \right] \! .~
\end{equation}
Then, from \eq{eq4_24} the following expression for the variance is obtained,
\begin{eqnarray}
  \!\!\!\!\!\!\!\!&&\sigma(\tau)^2 
  = \sigma_0^2
  + \left( \frac{\hbar}{2 M \sigma_0 \gamma} \right)^2
  \!\left[ 1 - \exp(-\gamma\tau) \right]^2
  \nonumber \\ 
  \label{eq5_7}
  \!\!\!\!\!\!\!\!&&+\frac{2}{M\beta\gamma^2} 
  \! \left[   \gamma\tau - \frac{3}{2} + 2 \exp(-\gamma\tau) - \frac{1}{2}  \exp(-2\gamma\tau) \right] 
  \! .
\end{eqnarray}
The third term on the right hand side coincides with the classical mean square displacement
obtained from the Langevin equation \req{eq5_3}.
For large values of $\tau$ one recovers the diffusive law $\sigma(\tau) \approx 2D\tau$,
where $D = 1/M\beta\gamma$ is the diffusion coefficient.

These results summarize the behavior of the classical particle and of the corresponding quantum particle
obtained when the correlated initial conditions described in \secb{sec3} are used.
What is the behavior which is obtained when factorized uncorrelated initial conditions of the form of \eq{eq2} are used?
Without repeating the calculations, the results for uncorrelated initial conditions can be obtained
setting $X_a = 0$ in \eq{eq3_2}.
Proceeding in a similar way, an effective action $A'[X,\xi] = A[X,\xi] + \Delta A[X,\xi]$ is obtained,
which differs from the effective action $A[X,\xi]$ given by \eq{eq4_11} for the additional term
\begin{eqnarray}
  \label{eq5_8}
  \Delta A[X,\xi] 
  = \int_{t_a}^{t_b} dt \, \xi(t) K(t-t_a) X_a  \, .
\end{eqnarray}
In the white noise limit, it reduces to $\Delta A[X,\xi] \approx M \gamma X_a \xi_a$.
Proceeding in a similar way, one can now compute the corresponding probability density at a generic time $\tau = t_b - t_a$. 
The result is again a Gaussian probability density of the form of \eq{eq4_22},
but the average position and mean square displacement are now given by
\begin{eqnarray}
  \label{eq5_9}
  &&X_\mathrm{av}(\tau) = x_0 \exp(-\gamma\tau) + \frac{v_0}{\gamma} \left[ 1 - \exp(-\gamma\tau)\right] \, ,
  \\ 
  &&\sigma(\tau)^2 \!= \sigma_0^2\exp(-2\gamma\tau) 
  \!+\! \left( \! \frac{\hbar}{2 M \sigma_0 \gamma} \! \right)^2 \!\! \left[ 1 - \exp(-\gamma\tau) \right]^2
  \nonumber \\ 
  \label{eq5_10}
  &&+\frac{2}{M\beta\gamma^2} 
  \! \left[   \gamma\tau - \frac{3}{2} + 2 \exp(-\gamma\tau) - \frac{1}{2}  \exp(-2\gamma\tau) \right] \! . ~~
\end{eqnarray}
These expressions differ at any $\tau > 0$ from those obtained starting from the correlated initial conditions:
respect to the average position defined by \eqs{eq4_5} and \req{eq5_4},
in the $X_\mathrm{av}(\tau)$ given by \eq{eq5_9} above the initial coordinate $x_0$
is multiplied by a factor $\exp(-\gamma\tau)$ and thus it \emph{does not affect} the asymptotic average position of the particle, which turns out to be
$X_\mathrm{av}(\tau \gg \gamma^{-1}) = v_0/\gamma$, instead of $x_0 + v_0/\gamma$.
This result has no physical sense, since, e.g., the overall distance covered by a classical free Brownian particle
(and by the center of mass of the corresponding quantum wave packet) would be frame-dependent,
\begin{eqnarray}
  \label{eq5_10b}
  L = X_\mathrm{av}(+\infty) - X_\mathrm{av}(t_0) =  \frac{v_0}{\gamma} - x_0 \, .
\end{eqnarray}
Even more surprisingly, in the mean square displacement given by \eq{eq5_10},
with respect to that in \eq{eq5_7}, the contribution $\sigma_0^2$ coming
from the initial quantum uncertainty appears multiplied by a damping factor $\exp(-2\gamma\tau)$
and goes to zero asymptotically.
It follows that for $\tau \gg \gamma^{-1}$ no contribution from the initial uncertainty
would affect the mean square displacement of the particle, however large $\sigma_0$ may be.
This is also in disagreement with the corresponding classical formula.

The strange behavior predicted by \eqs{eq5_9} and \req{eq5_10} is only due 
to the form of the uncorrelated initial conditions, i.e., to the hypothesis of absence of correlation.
between central particle and environment in the initial state.
This is best shown by the analogous effect which takes place in the classical model.
The classical uncorrelated initial conditions are obtained by assuming that at $t = t_0$ 
the classical oscillator coordinates have canonical distributions with average coordinate 
$\langle q_{n0} \rangle = 0$.
In this case, in order for the fluctuation-dissipation theorem to hold,
i.e., $\langle R(t) \rangle = 0$ and $\beta \langle R(t) R(s) \rangle = K(t-s)$,
the generalized Langevin equation \req{eq2_2} must be rewritten as~\cRef{Ford1987a}
\begin{equation}
  \label{eq5_11}
  M \ddot{x} + V'(x,t) = - \int_{t_0}^t ds \, K(t-s) \dot{x}(s) + \Delta F(t) + R(t) \, .
\end{equation}
Here the random force $R'(t)$, given by \eq{eq2_4}, has been split as $R(t) = R'(t) + \Delta F(t)$,
with 
\begin{eqnarray}
  \label{eq5_12}
  \!\!\!\!\!\!\!\!&&\Delta F(\tau) = - \sum_n m \omega_n^2 x_0 \cos \left(\omega_n \tau \right) \equiv - K(t-t_0) x_0 \, ,~
  \\
  \label{eq5_13}
  \!\!\!\!\!\!\!\!&&R'(t) = \sum_n m \omega_n^2 \left\{ 
  \vphantom{\frac{\dot{q}}{\omega}}
  q_{n0} \cos \left[\omega_n (t - t_0) \right] \right. 
  \nonumber \\
  \!\!\!\!\!\!\!\!&&~~~~~~~~~~~~
  + \left. \frac{(\dot{q}_{n0} - \dot{x}_0)}{\omega_n} \sin \left[\omega_n (t - t_0) \right] \right\},~~
\end{eqnarray}
It is to be noticed that the new random force $R'(t)$ is not translation invariant and that
the force term $\Delta F(t)$ is not stochastic.
How does $\Delta F(t)$ modify the classical average trajectory?
In the white noise limit $\Delta F(t)$ is just an initial bump, i.e.,
$\Delta F(t) = -2x_0 M \gamma \delta(t-t_0)$.
However, due to the delta function, 
the Brownian particle receives a finite momentum proportional to $x_0$,
\begin{eqnarray}
  \label{eq5_13b}
\Delta p = \int dt \, F(t) = - M \gamma x_0 \, . 
\end{eqnarray}
This is equivalent to an effective (frame-dependent) initial velocity
\begin{eqnarray}
  \label{eq5_13c}
  v_0' = v_0 - \frac{\Delta p}{M} = v_0 - \gamma x_0 \, .
\end{eqnarray}
With such an initial velocity, the solution for the classical average position
becomes equal to the frame-dependent solution in \eq{eq5_9}.

Thus, uncorrelated initial conditions give rise to a spurious force term,
which strongly influences the average position of the quantum wave packet and of the classical particle
in the same way.
The additional force term $\Delta F(t)$ of the classical problem, given by \eq{eq5_12},
corresponds to the additional action term $\Delta A[X,\xi]$ of the quantum problem, given by \eq{eq5_8}.

\section{Conclusion}
\label{conclusion}
%
In the present study, the problem of the consistent formulation of the initial conditions in the oscillator model of quantum linear dissipative systems was considered.
On the analogy of the classical model,
a novel simple form of initial conditions was obtained,
in which the total density matrix factorizes as the product of the density matrices of the subsystems, i.e. of the central systems and the environment.
However, there is a correlation between central system and environment,
arising from assuming thermal equilibrium of the environment at the initial time,
which is properly taken into account by the form of the environment density matrix.

As a check on the new form of such factorized --- but correlated --- initial conditions,
the dynamics of a forced Brownian particle with arbitrary colored noise was studied.
Starting from correlated initial conditions,
it was shown that a quantum wave packet moves like the corresponding classical Brownian particle,
i.e., with the same average position.
In the white noise and large temperature limit, also the mean square displacement reduces to its classical counterpart. 

On the contrary, both the average position and the mean square displacement are modified in a unphysical way by a spurious force term appearing if uncorrelated initial conditions are assumed -- the average positions becomes frame-dependent and the mean square displacement looses memory of the initial uncertainty. 
Similar effects take place in the classical and quantum model.
This results demonstrates that the so-called \emph{uncorrelated initial conditions} actually represent an environment in a state very far from thermal equilibrium.

The whole issue about the form of the initial conditions, as well as that of the Lagrangian, can be summarized from the point of view of the general underlying symmetries:

(a) Translation and reflection invariance are required symmetries for the environment oscillator sector of the Lagrangian, describing the interaction between central system and environment, in order to re-obtain the Langevin equation.

(b) Consistently, the same symmetries have to be present in the initial conditions of the environment oscillators, at least in the applications of the Feynman-Vernon model to classical and quantum Brownian motion.

(c) The points above are not sufficient for a complete specification of the initial conditions of the environment oscillators in the quantum problem and one more prescription is needed: the environment initial conditions must be invariant under time-reversal, in order to represent a state of thermal equilibrium.


\begin{acknowledgments}
This revision of Ref.~\cite{Patriarca1996a} was made possible by the support of the European Regional Development Fund (ERDF) Center of Excellence (CoE) program grant TK133 and the Estonian Research Council through Institutional Research Funding Grants (IUT) No. IUT-39-1, IUT23-6, and Personal Research Funding Grant (PUT) No. PUT-1356. 
\end{acknowledgments}

\newpage
\onecolumngrid



\end{document}